\pdfoutput=1 

\documentclass{article}%
\usepackage{amsfonts}
\usepackage{amssymb}
\usepackage{amsmath}
\usepackage{graphicx}%
\providecommand{\U}[1]{\protect\rule{.1in}{.1in}}

\begin{document}

\title{A Very Common Fallacy in Quantum Mechanics:\\Superposition, Delayed Choice, Quantum Erasers, Retrocausality, and All That}
\author{David Ellerman\\University of California at Riverside}
\maketitle

\begin{abstract}
There is a very common fallacy, here called the \textit{separation fallacy},
that is involved in the interpretation of quantum experiments involving a
certain type of separation such as the: double-slit experiments, which-way
interferometer experiments, polarization analyzer experiments, Stern-Gerlach
experiments, and quantum eraser experiments. It is the separation fallacy that
leads not only to flawed textbook accounts of these experiments but to flawed
inferences about retrocausality in the context of "delayed choice" versions of
separation experiments.

\end{abstract}
\tableofcontents

\section{Introduction: Separation Fallacy}

There is a very common fallacy, here called the \textit{separation fallacy},
that is involved in the interpretation of quantum experiments involving a
certain type of separation such as the:

\begin{itemize}
\item double-slit experiments,

\item which-way interferometer experiments,

\item polarization analyzer experiments,

\item Stern-Gerlach experiments, and

\item quantum eraser experiments.
\end{itemize}

In each case, given an incoming quantum particle, the apparatus creates a
certain labelled or tagged (i.e., entangled) superposition of certain
eigenstates (the "separation"). Detectors can be placed in certain positions
(determined by the tags) so that when the evolving superposition state is
finally projected or collapsed by the detectors, then only one of the
eigenstates can register at each detector. The \textit{separation fallacy}
mistakes the creation of a tagged or entangled superposition for a
measurement. Thus it treats the particle as if it had already been projected
or collapsed to an eigenstate at the separation apparatus rather than at the
later detectors. But if the detectors were suddenly removed while the particle
was in the apparatus, then the superposition would continue to evolve and have
distinctive effects (e.g., interference patterns in the two-slit experiment).

Hence the separation fallacy makes it \textit{seem} that by the delayed choice
to insert or remove the appropriately positioned detectors, one can
\textit{retro-cause} either a collapse to an eigenstate or not at the
particle's entrance into the separation apparatus.

The separation fallacy is remedied by:

\begin{itemize}
\item taking superposition seriously, i.e., by seeing that the separation
apparatus created an entangled \textit{superposition} state of the
alternatives (regardless of what happens later) which evolves until a
measurement is taken, and

\item taking into account the role of detector placement ("contextuality"),
i.e., by seeing that if a suitably positioned detector, as determined by the
positional labels or tags, can only detect one collapsed eigenstate, then it
does not mean that the particle was \textit{already} in that eigenstate prior
to the measurement (e.g., it does not mean that the particle "went through one
slit," "took one arm," or was already in a polarization or spin eigenstate).
\end{itemize}

The separation fallacy will be first illustrated in a non-technical manner for
the first four experiments. Then the lessons will be applied in a slightly
more technical discussion of quantum eraser experiments--where, due to the
separation fallacy, incorrect inferences about retrocausality have been rampant.

\section{Double-slit experiments}

In the well-known setup for the double-slit experiment, if a detector $D_{1}$
is placed a small finite distance after slit 1 so a particle "going through
the other slit" cannot reach the detector, then a hit at the detector is
usually interpreted as "the particle went through slit 1."%

\begin{center}
\includegraphics[
natheight=1.187500in,
natwidth=1.161500in,
height=1.2453in,
width=1.2177in
]%
{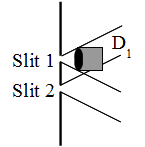}%
\\
Figure 1a
\end{center}

But this is incorrect. The particle is in a superposition state, which we
might represent schematically as $\left\vert Slit1\right\rangle +\left\vert
Slit2\right\rangle $, that evolves until it hits the detector which projects
(or collapses) the superposition to one of (the evolved versions of) the
slit-eigenstates. The particle's state was not collapsed earlier so it was not
previously in the $\left\vert Slit1\right\rangle $ eigenstate, i.e., it did
not "go through slit 1."

Thus what is called "detecting which slit the particle went through" is a
misinterpretation. It is only placing a detector in such a position so that
when the superposition projects to an eigenstate, only one of the eigenstates
can register in that detector. It is about \textit{detector placement}; it is
not about which-slit.

By erroneously talking about the detector "showing the particle went through
slit 1," we imply a type of retro-causality. If the detector is suddenly
removed after the particle has passed the slits, then the superposition state
continues to evolve and shows interference on the far wall (not shown)---in
which case people say "the particle went through both slits." Thus the "bad
talk" makes it seem that by removing or inserting the detector after the
particle is beyond the slits, one can retro-cause the particle to go through
both slits or one slit only.

This sudden removal or insertion of detectors that can only detect one of the
slit-eigenstates is a version of Wheeler's delayed choice thought-experiment
\cite{Wheeler:delayed}.%

\begin{center}
\includegraphics[
natheight=1.850000in,
natwidth=4.832900in,
height=1.9246in,
width=4.9813in
]%
{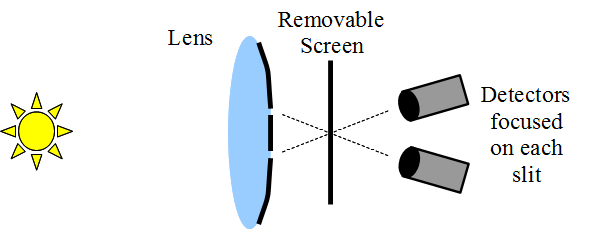}%
\\
Figure 1b. Wheeler's Delayed Choice 2 Slit Experiment
\end{center}

In Wheeler's version of the experiment, there are two detectors which are
positioned behind the removable screen so they can only detect one of the
projected (evolved) slit eigenstates when the screen is removed. The choice to
remove the screen or not is delayed until after a photon has traversed the two slits.

\begin{quotation}
"In the one case [screen in place] the quantum will ... contribute to the
record of a two-slit interference fringe. In the other case [screen removed]
one of the two counters will go off and signal in which beam--and therefore
from which slit--the photon has arrived." \cite[p. 13]{Wheeler:delayed}
\end{quotation}

The separation fallacy is involved when Wheeler infers from the fact that one
of the specially-placed detectors went off that the photon had come from one
of the slits--as if there had been a projection to one of the slit eigenstates
at the slits rather than later at the detectors.

Such descriptions using the separation fallacy are unfortunately common and
have generated a spate of speculations about retrocausality.

\section{Which-way interferometer experiments}

Consider a Mach-Zehnder-style interferometer with only one beam-splitter
(e.g., half-silvered mirror) at the photon source which creates the photon
superposition: $\left\vert T1\right\rangle +\left\vert R1\right\rangle $
(which stand for "Transmit" to the upper arm or "Reflect" into the lower arm
at the first beam-splitter).%

\begin{center}
\includegraphics[
natheight=2.529200in,
natwidth=3.733500in,
height=2.6207in,
width=3.8542in
]%
{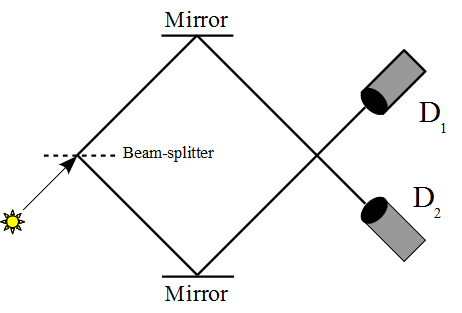}%
\\
Figure 2: One beam-splitter
\end{center}

When detector $D_{1}$ registers a hit, it is said that "the photon was
reflected and thus took the lower arm" of the interferometer and similarly for
$D_{2}$ and passing through into the upper arm. This is the interferometer
analogue of putting two up-close detectors after the two slits in the two-slit experiment.

And this standard description is incorrect for the same reasons. The photon
stays in the superposition state until the detectors force a projection to one
of the (evolved) eigenstates. If the projection is to the evolved $\left\vert
R1\right\rangle $ eigenstate then only $D_{1}$ will get a hit, and similarly
for $D_{2}$ and the evolved version of $\left\vert T1\right\rangle $. The
point is that the placement of the detectors (like in the double-slit
experiment) only captures one or the other of the projected eigenstates.

Now insert a second beam-splitter as in the following diagram.%

\begin{center}
\includegraphics[
natheight=2.529200in,
natwidth=3.733500in,
height=2.6207in,
width=3.8542in
]%
{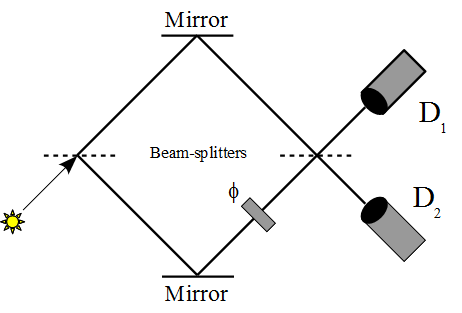}%
\\
Figure 3: Two beam-splitters
\end{center}

It is said that the second beam-splitter "erases" the "which-way information"
so that a hit at either detector could have come from either arm, and thus an
interference pattern emerges.

But this is also incorrect. The superposition state $\left\vert
T1\right\rangle +\left\vert R1\right\rangle $ (which contains no which-way
information) is further transformed at the second beam-splitter to the
superposition $\left\vert T1,T2\right\rangle +\left\vert T1,R2\right\rangle
+\left\vert R1,T2\right\rangle +\left\vert R1,R2\right\rangle $ that can be
regrouped according to what can register at each detector:

\begin{center}
$\left[  \left\vert T1,R2\right\rangle +\left\vert R1,T2\right\rangle \right]
_{D_{1}}+\left[  \left\vert T1,T2\right\rangle +\left\vert R1,R2\right\rangle
\right]  _{D_{2}}$.
\end{center}

The so-called "which-way information" was not there to be "erased" since the
particle did not take one way or the other in the first place. The second
beam-splitter only allows the superposition state $\left[  \left\vert
T1,R2\right\rangle +\left\vert R1,T2\right\rangle \right]  _{D_{1}}$to be
registered at $D_{1}$ or the superposition state $\left[  \left\vert
T1,T2\right\rangle +\left\vert R1,R2\right\rangle \right]  _{D_{2}}$ to be
registered at $D_{2}$. By using a phase shifter $\phi$, an interference
pattern can be recorded at each detector since each one is now detecting a
superposition that will involve interference.

By inserting or removing the second beam-splitter after the particle has
traversed the first beam-splitter (as in \cite{Wheeler:delayed}), the
separation fallacy makes it seem that we can retro-cause the particle to go
through both arms or only one arm.

The point is not the second beam-splitter but the detectors being able to
register the collapse to either eigenstate and thus the interference between
them. Instead of inserting the second beam-splitter, we could rig up more
mirrors, a lens, and a single detector so that when the single detector causes
the collapse, then it is will register either arm-eigenstate.%

\begin{center}
\includegraphics[
natheight=3.151500in,
natwidth=5.753700in,
height=3.2588in,
width=5.9256in
]%
{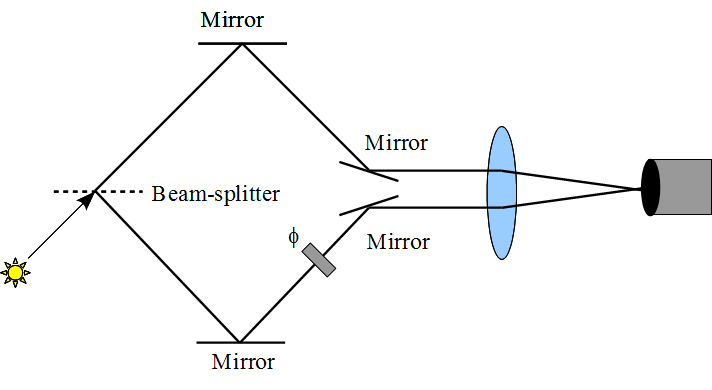}%
\\
Figure 4: Detector placed to register all hits
\end{center}

This might also be (mis)interpreted as "erasing" the "which-way information"
but in fact the photon did not go through just one arm so there was no such
information to be erased. The point is the positioning of the detector so that
it detects the evolved superposition $\left\vert T1\right\rangle +\left\vert
R1\right\rangle $ that will show interference. Any setup that would allow a
detector to register both collapsed eigenstates (and thus to register the
interference effects of the evolving superposition) would \textit{ipso facto}
be a setup that could be (mis)interpreted as "erasing" the "which-way
information." That is why the separation fallacy is so persistent in the
interpretation of which-way interferometer and other quantum separation experiments.

\section{Polarization analyzers and loops}

Another common textbook example of the separation fallacy is the treatment of
polarization analyzers such as calcite crystals that are said to create two
orthogonally polarized beams in the upper and lower channels, say $\left\vert
v\right\rangle $ and $\left\vert h\right\rangle $ from an arbitrary incident beam.%

\begin{center}
\includegraphics[
natheight=0.937600in,
natwidth=5.955800in,
height=0.956in,
width=5.0207in
]%
{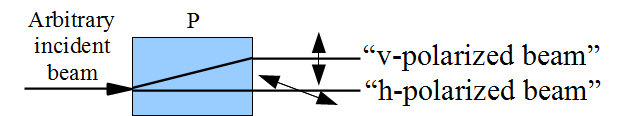}%
\\
Figure 5: $vh$-analyzer
\end{center}

The output from the analyzer $P$ is routinely described as a "vertically
polarized" beam and "horizontally polarized" beam as if the analyzer was
itself a measurement that collapsed or projected the incident beam to either
of those polarization eigenstates. This \textit{seems} to follow because if
one positions a detector in the upper beam then only vertically polarized
photons are observed and similarly for the lower beam and horizontally
polarized photons. A blocking mask in one of the beams has the same effect as
a detector to project the photons to eigenstates. If a blocking mask in
inserted in the lower beam, then only vertically polarized photons will be
found in the upper beam, and vice-versa.

But here again, the story is about detector (or blocking mask) placement
("placement" is more precise than "contextuality"); it is not about the
analyzer supposedly projecting a photon into one or the other of the
eigenstates. The analyzer puts the incident photons into a superposition
state, an entangled superposition state that associates polarization and the
spatial channel. If a detector is placed in, say, the upper channel, then
\textit{that} is the measurement that collapses the evolved superposition
state. If the collapse is to the vertical polarization eigenstate then it will
register only in the upper detector and similarly for a collapse to the
horizontal polarization eigenstate for any detector placed in the lower
channel. Thus it is misleadingly said that the upper beam was already
vertically polarized and the lower beam was already horizontally polarized as
if the analyzer had already done the projection to one of the eigenstates.

If the analyzer had in fact induced a collapse to the eigenstates, then any
prior polarization of the incident beam would be lost. Hence assume that the
incident beam was prepared in a specific polarization of, say, $\left\vert
45^{\circ}\right\rangle $ half-way between the states of vertical and
horizontal polarization. Then follow the $vh$-analyzer $P$ with its inverse
$P^{-1}$ to form an analyzer loop \cite{French:qp}.%

\begin{center}
\includegraphics[
natheight=0.921600in,
natwidth=3.822400in,
height=0.9719in,
width=3.9465in
]%
{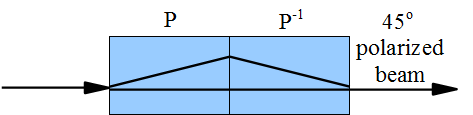}%
\\
Figure 6: $vh$-analyzer loop
\end{center}

The characteristic feature of an analyzer loop is that it outputs the same
polarization, in this case $\left\vert 45^{\circ}\right\rangle $, as the
incident beam. This would be impossible if the $P$ analyzer had in fact
rendered all the photons into a vertical or horizontal eigenstate thereby
destroying the information about the polarization of the incident beam. But
since no collapsing measurement was in fact made in $P$ or its inverse, the
original beam can be the output of an analyzer loop.

Very few textbooks realize there is even a problem with presenting a
polarization analyzer such as a calcite crystal as creating two beams with
orthogonal eigenstate polarizations---rather than creating a tagged
superposition state so that appropriately positioned detectors can detect only
one eigenstate when the detectors cause the projections to eigenstates.

One (partial) exception is Dicke and Wittke's text \cite{Dicke:qm}. At first
they present polarization analyzers as if they measured polarization and thus
"destroyed completely any information that we had about the polarization"
\cite[p. 118]{Dicke:qm} of the incident beam. But then they note a problem:

\begin{quotation}
"The equipment [polarization analyzers] has been described in terms of devices
which measure the polarization of a photon. Strictly speaking, this is not
quite accurate." \cite[p. 118]{Dicke:qm}
\end{quotation}

They then go on to consider the inverse analyzer $P^{-1}$ which combined with
$P$ will form an analyzer loop that just transmits the incident photon unchanged.

They have some trouble squaring this with their prior statement about the $P$
analyzer destroying the polarization of the incident beam but they, unlike
most texts, struggle with getting it right.

\begin{quotation}
"Stating it another way, although [when considered by itself] the polarization
$P$ completely destroyed the previous polarization $Q$ [of the incident beam],
making it impossible to predict the result of the outcome of a subsequent
measurement of $Q$, in [the analyzer loop] the disturbance of the polarization
which was effected by the box $P$ is seen to be revocable: if the box $P$ is
combined with another box of the right type, the combination can be such as to
leave the polarization $Q$ unaffected." \cite[p. 119]{Dicke:qm}
\end{quotation}

They then go on to correctly note that the polarization analyzer $P$ did not
in fact project the incident photons into polarization eigenstates.

\begin{quotation}
"However, it should be noted that in this particular case [sic!], the first
box $P$ in [the first half of the analyzer loop] did not really measure the
polarization of the photon: no determination was made of the channel ... which
the photon followed in leaving the box $P$." \cite[p. 119]{Dicke:qm}
\end{quotation}

There is some classical imagery (like Schr\"{o}dinger's cat running around one
side or the other side of a tree) that is sometimes used to illustrate quantum
separation experiments when in fact it only illustrates how classical imagery
can be misleading. Suppose an interstate highway separates at a city into both
northern and southern bypass routes--like the two channels in a polarization
analyzer loop. One can observe the bypass routes while a car is in transit and
find that it is in one bypass route or another. But after the car transits
whichever bypass it took without being observed and rejoins the undivided
interstate, then it is said that the which-way information is erased so an
observation cannot elicit that information.

This is not a correct description of the corresponding quantum separation
experiment since the classical imagery does not contemplate superposition
states. The particle-as-car is in a tagged superposition of the two routes
until an observation (e.g., a detector or "road block") collapses the
superposition to one eigenstate or the other. Correct descriptions of quantum
separation experiments require taking superposition seriously--so classical
imagery should only be used \textit{cum grano salis}.

This analysis might be rendered in a more technical but highly schematic way.
The photons in the incident beam have a particular polarization $\left\vert
\psi\right\rangle $ such as $\left\vert 45^{\circ}\right\rangle $ in the above
example. This polarization state can be represented or resolved in terms of
the $vh$-basis as:

\begin{center}
$\left\vert \psi\right\rangle =\left\langle v|\psi\right\rangle \left\vert
v\right\rangle +\left\langle h|\psi\right\rangle \left\vert h\right\rangle $.
\end{center}

The effect of the $vh$-analyzer $P$ might be represented as tagging the
vertical and horizontal polarization states with the upper and lower (or
straight) channels so the $vh$-analyzer puts an incident photon into the
\textit{superposition} state:

\begin{center}
$\left\langle v|\psi\right\rangle \left\vert v\right\rangle _{U}+\left\langle
h|\psi\right\rangle \left\vert h\right\rangle _{L}$,
\end{center}

\noindent\textit{not} into an eigenstate of $\left\vert v\right\rangle $ in
the upper channel or an eigenstate $\left\vert h\right\rangle $ in the lower channel.

If a blocker or detector were inserted in either channel, then this
superposition state would project to one of the eigenstates, and then (as
indicated by the tags) only vertically polarized photons would be found in the
upper channel and horizontally polarized photons in the lower channel.

The separation fallacy is to describe the $vh$-analyzer \textit{as if} the
analyzer's effect by itself was to project an incident photon either into
$\left\vert v\right\rangle $ in the upper channel or $\left\vert
h\right\rangle $ in the lower channel--instead of only creating the above
tagged superposition state. The mistake of describing the unmeasured
polarization analyzer as creating two beams of eigenstate polarized photons is
analogous to the mistake of describing a particle as going through one slit or
the other in the unmeasured-at-slits double-slit experiment--and similarly for
the other separation experiments.

It is fallacious to reason that "we know the photons are in one polarization
state in one channel and in the orthogonal polarization state in the other
channel \textit{because} that is what we find when we measure the channels,"
just as it is fallacious to reason "the particle has to go through one slit or
another (or one arm or another in the interferometer experiment)
\textit{because} that is what we find when we measure it." This purely
\textit{operational} (or "Copenhagen") description does not take superposition
seriously since a superposition state is not "what we find when we measure."

In the analyzer \textit{loop}, no measurement (detector or blocker) is made
after the $vh$-analyzer. It is followed by the inverse $vh$-analyzer $P^{-1}$
which has the inverse effect of removing the $U$ and $L$ tags from the
superposition state $\left\langle v|\psi\right\rangle \left\vert
v\right\rangle _{U}+\left\langle h|\psi\right\rangle \left\vert h\right\rangle
_{L}$ so that a photon exits the loop in the untagged superposition state:

\begin{center}
$\left\langle v|\psi\right\rangle \left\vert v\right\rangle +\left\langle
h|\psi\right\rangle \left\vert h\right\rangle =\left\vert \psi\right\rangle $.
\end{center}

\noindent The inverse $vh$-analyzer does not "erase" the which-polarization
information since there was no measurement--to reduce the superposition state
to eigenstate polarizations in the channels of the analyzer loop--in the first
place. The inverse $vh$-analyzer \textit{does erase} the which-channel tags so
the original state $\left\langle v|\psi\right\rangle \left\vert v\right\rangle
+\left\langle h|\psi\right\rangle \left\vert h\right\rangle =\left\vert
\psi\right\rangle $ is restored (which could be viewed as a type of
interference effect, e.g., \cite[Sections 7-4, 7-5]{French:qp}).

\section{Stern-Gerlach experiment}

We have seen the separation fallacy in the standard treatments of the
double-slit experiment, which-way interferometer experiments, and in
polarization analyzers. In spite of the differences between those separation
experiments, there was that common (mis)interpretative theme. Since the
"logic" of the polarization analyzers is followed in the Stern-Gerlach
experiment (with spin playing the role of polarization), it is not surprising
that the same fallacy occurs there.%

\begin{center}
\includegraphics[
natheight=1.563200in,
natwidth=3.393800in,
height=1.6286in,
width=3.507in
]%
{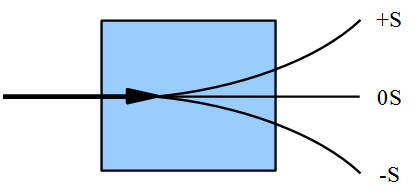}%
\\
Figure 7: Stern-Gerlach Apparatus
\end{center}

And again, the fallacy is revealed by considering the Stern-Gerlach analogue
of an analyzer loop. One of the very few texts to consider such a
Stern-Gerlach analyzer loop is \textit{The Feynman Lectures on Physics:
Quantum Mechanics (Vol. III)} where it is called a "modified Stern-Gerlach
apparatus" \cite[p. 5-2]{Feynman:lect3}.%

\begin{center}
\includegraphics[
natheight=1.721700in,
natwidth=5.025800in,
height=1.7913in,
width=5.1801in
]%
{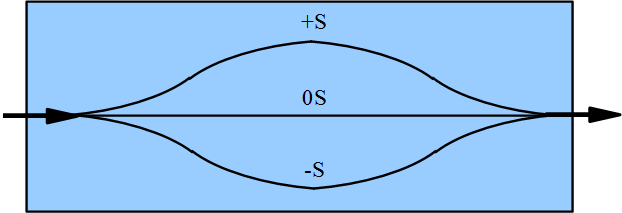}%
\\
Figure 8: Stern-Gerlach Loop
\end{center}

Ordinarily texts represent the Stern-Gerlach apparatus as separating particles
into spin eigenstates denoted by, say, $+S,0S,-S$. But as in our other
examples, the apparatus does not project the particles to eigenstates. Instead
it creates a superposition state so that with a detector in a certain
position, then as the detector causes the collapse to a spin eigenstate, the
detector will only see particles of one spin state. Alternatively if the
collapse is caused by placing blocking masks over two of the beams, then the
particles in the third beam will all be those that have collapsed to the same
eigenstate. It is the detectors or blockers that cause the collapse or
projection to eigenstates, not the prior separation apparatus.

We previously saw how a polarization analyzer, contrary to the statement in
many texts, does not lose the polarization information of the incident beam
when it "separates" the beam (into a positionally-tagged superposition state).
In the context of the Stern-Gerlach apparatus, Feynman similarly remarks:

\begin{quotation}
"Some people would say that in the filtering by T we have 'lost the
information' about the previous state ($+S$) because we have 'disturbed' the
atoms when we separated them into three beams in the apparatus T. But that is
not true. The past information is not lost by the \textit{separation} into
three beams, but by the \textit{blocking masks} that are put in\ldots."
\cite[p. 5-9 (italics in original)]{Feynman:lect3}
\end{quotation}

\section{The Separation Fallacy}

We have seen the same fallacy of interpretation in two-slit experiments,
which-way interferometer experiments, polarization analyzers, and
Stern-Gerlach experiments. The common element in all the cases is that there
is some 'separation' apparatus that puts a particle into a certain
superposition of eigenstates in such a manner that when an appropriately
positioned detector induces a collapse to an eigenstate, then the detector
will only register one of the eigenstates. The separation fallacy is that this
is misinterpreted as showing that the particle was already in that eigenstate
in that position as a result of the previous 'separation.' The quantum erasers
are elaborated versions of these simpler experiments, and a similar separation
fallacy arises in that context.

\section{One photon quantum eraser experiment}

A simple quantum eraser can be devised using a single beam of photons as in
\cite{HK:DIYQeraser}. We start with the standard two-slit setup.%

\begin{center}
\includegraphics[
natheight=2.594700in,
natwidth=3.700800in,
height=2.6861in,
width=3.8215in
]%
{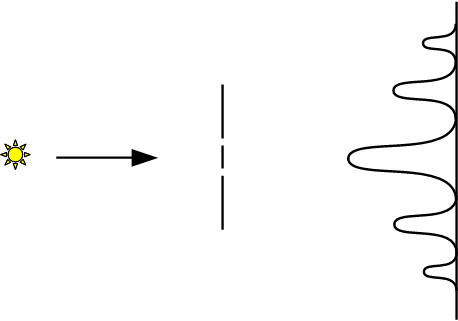}%
\\
Figure 9: Two-slit setup
\end{center}

After the two slits, a photon could be schematically represented as being in a
superposition state $\left\vert s1\right\rangle +\left\vert s2\right\rangle $
(where $s1$ and $s2$ stand for the two slits) which evolves with interference
to give the familiar pattern on the far wall.

Then a horizontal polarizer is place in front of slit 1 and a vertical
polarizer in front of slit 2.%

\begin{center}
\includegraphics[
natheight=2.594700in,
natwidth=3.700800in,
height=2.6861in,
width=3.8215in
]%
{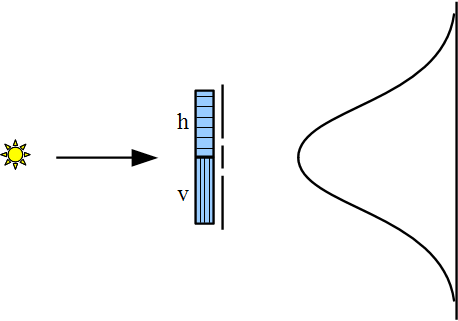}%
\\
Figure 10: Two-slit setup with $v,h$ polarizers
\end{center}

After the two slits, a photon is in a state that entangles the spatial slit
states and the polarization states which might be represented as: $\left\vert
s1\right\rangle \otimes\left\vert h\right\rangle +\left\vert s2\right\rangle
\otimes\left\vert v\right\rangle $ (for a discussion of this type of
entanglement, see \cite{kwiat:2photon}). But as this superposition evolves, it
cannot be separated into a superposition of the slit-states as before, so the
interference disappears.

Then a $+45^{o}$ polarizer is inserted between the two-slit screen and the
wall. This transforms the evolving state to:

\begin{center}
$\left\vert s1\right\rangle \otimes\left\vert 45^{o}\right\rangle +\left\vert
s2\right\rangle \otimes\left\vert 45^{o}\right\rangle =\left[  \left\vert
s1\right\rangle +\left\vert s2\right\rangle \right]  \otimes\left\vert
45^{o}\right\rangle $
\end{center}

\noindent so that the $\left\vert s1\right\rangle +\left\vert s2\right\rangle
$ term will show interference in a "fringe" pattern when the $45^{o}$
polarized photons hit the wall. If we had inserted a $-45^{o}$ polarizer, then
again interference in an "antifringe" pattern would appear as the $-45^{o}$
polarized photons hit the wall. The sum of the fringe and antifringe patterns
gives the no-interference pattern of the previous figure.%

\begin{center}
\includegraphics[
natheight=2.594700in,
natwidth=3.700800in,
height=2.6861in,
width=3.8215in
]%
{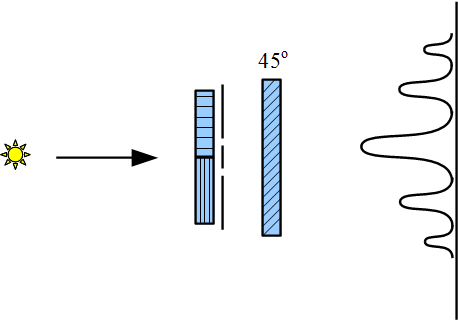}%
\\
Figure 11: $+45^{o}$ polarizer and fringe pattern
\end{center}

A common description of this type of quantum eraser experiment is that the
insertion of the $h,v$ polarizers "marks" the photons with "which-slit
information" (Figure 10) that destroys the interference--even if the
horizontal or vertical polarization is not measured at the wall. If the
horizontal or vertical polarization was measured at the wall, then the evolved
superposition state $\left\vert s1\right\rangle \otimes\left\vert
h\right\rangle +\left\vert s2\right\rangle \otimes\left\vert v\right\rangle $
would collapse to the evolved version of $\left\vert s1\right\rangle $ (if $h$
was found) or $\left\vert s2\right\rangle $ (if $v$ was found). This is said
to reveal the so-called "which-slit information" that the photon went through
slit 1 or slit 2, i.e., that at the slits, the photon was already in the state
$\left\vert s1\right\rangle $ or $\left\vert s2\right\rangle $ instead of
being in the entangled superposition state. By incorrectly inferring that the
photon was in one state or the other at the slits--while it would have to "go
through both slits" to yield the interference pattern obtained by inserting
the $45^{o}$ polarizer--we seem to be able to retrocause the particle to go
through one slit or both slits by withdrawing or inserting the $45^{o}$
polarizer after a photon has traversed the two slits.\ 

It is precisely the separation fallacy that leads to this inference of
retrocausality. In the situation of Figure 10, the photon superposition state
$\left\vert s1\right\rangle \otimes\left\vert h\right\rangle +\left\vert
s2\right\rangle \otimes\left\vert v\right\rangle $ evolves until it hits the
wall. The slit states are indeed marked, tagged, labelled, or entangled with
polarization states but this is incorrectly called "which-way information" as
if it could "reveal" that the photon was in the state $\left\vert
s1\right\rangle $ or $\left\vert s2\right\rangle $ at the slits, i.e., that it
went through slit 1 or slit 2.

Also it might be noted that the insertion of a $+45^{o}$ or $-45^{o}$
polarizer does not "restore" the original interference pattern of Figure 9 but
picks out the fringe or antifringe interference patterns out of the Figure 10
"mush" of hits.

\section{Two photon quantum eraser experiment}

We now turn to one of the more elaborate quantum eraser experiments
\cite{Walborn:eraser}.%

\begin{center}
\includegraphics[
natheight=2.787500in,
natwidth=2.666800in,
height=2.8152in,
width=2.6936in
]%
{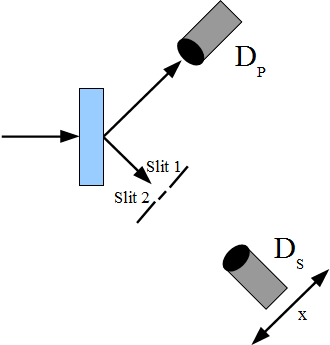}%
\\
Figure 12: Setup with two slits
\end{center}

A photon hits a down-converter which emits a "signal" $p$-photon entangled
with an "idler" $s$-photon with a superposition of orthogonal $\left\vert
x\right\rangle $ and $\left\vert y\right\rangle $ polarizations so the overall
state is:

\begin{center}
$\left\vert \Psi\right\rangle =\frac{1}{\sqrt{2}}\left[  \left\vert
x\right\rangle _{s}\otimes\left\vert y\right\rangle _{p}+\left\vert
y\right\rangle _{s}\otimes\left\vert x\right\rangle _{p}\right]  $.
\end{center}

The lower $s$-photon hits a double-slit screen, and will show an interference
pattern on the $D_{s}$ detector as the detector is moved along the $x$-axis.

Next two quarter-wave plates are inserted before the two-slit screen with the
fast axis of the one over slit $1$ oriented at $\left\vert +45^{\circ
}\right\rangle $ to the x-axis and the one over the slit $2$ with its fast
axis oriented at $\left\vert -45^{\circ}\right\rangle $ to the $x$-axis.%

\begin{center}
\includegraphics[
natheight=2.787500in,
natwidth=2.666800in,
height=2.8857in,
width=2.7607in
]%
{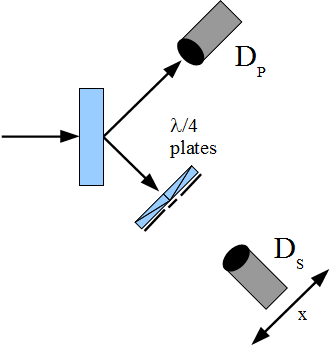}%
\\
Figure 13: Setup with $\lambda/4$ plates
\end{center}

Then Walborn et al. \cite{Walborn:eraser} give the overall state of the system
as (where the $s1$ and $s2$ tags refer to the two slits):

\begin{center}
$\left\vert \Psi\right\rangle =\frac{1}{2}\left[  \left(  \left\vert
L\right\rangle _{s1}\otimes\left\vert y\right\rangle _{p}+i\left\vert
R\right\rangle _{s1}\otimes\left\vert x\right\rangle _{p}\right)  +\left(
i\left\vert R\right\rangle _{s2}\otimes\left\vert y\right\rangle
_{p}-i\left\vert L\right\rangle _{s2}\otimes\left\vert x\right\rangle
_{p}\right)  \right]  $.
\end{center}

Then by measuring the linear polarization of the $p$-photon at $D_{p}$ and the
circular polarization at $D_{s}$, "which-slit information" is said to be
obtained and no interference pattern recorded at $D_{s}$.

For instance measuring $\left\vert x\right\rangle $ at $D_{p}$ and $\left\vert
L\right\rangle $ at $D_{s}$ imply $s2$, i.e., slit $2$. But as previously
explained, this does \textit{not} mean that the $s$-photon went through slit
$2$. It means we have positioned the two detectors \textit{in polarization
space}, say to measure $\left\vert x\right\rangle $ polarization at $D_{p}$
and $\left\vert L\right\rangle $ polarization at $D_{s}$, so only when the
superposition state collapses to $\left\vert x\right\rangle $ for the
$p$-photon and $\left\vert L\right\rangle $ for the $s$-photon do we get a hit
at both detectors.

This is the analogue of the one-beam-splitter interferometer where the
positioning of the detectors would only record one collapsed state which did
not imply the system was all along in that particular arm-eigenstate. The
phrase "which-slit" or "which-arm information" is a misnomer in that it
implies the system was already in a slit- or way-eigenstate and the so-called
measurement only revealed the information. Instead, it is only at the
measurement that there is a collapse or projection to an evolved
slit-eigenstate (not at the previous separation due to the two slits).

Walborn et al. indulge in the separation fallacy when they discuss what the
so-called "which-path information" reveals.

\begin{quotation}
Let us consider the first possibility [detecting $p$ before $s$]. If photon
$p$ is detected with polarization $x$ (say), then we know that photon $s$ has
polarization $y$ before hitting the $\lambda/4$ plates and the double slit. By
looking at [the above formula for $\left\vert \Psi\right\rangle $], it is
clear that detection of photon $s$ (after the double slit) with polarization
$R$ is compatible only with the passage of $s$ through slit $1$ and
polarization $L$ is compatible only with the passage of $s$ through slit $2$.
This can be verified experimentally. In the usual quantum mechanics language,
detection of photon $p$ before photon $s$ has prepared photon s in a certain
state. \cite[p. 4]{Walborn:eraser}
\end{quotation}

\noindent Firstly, the measurement that $p$ has polarization $x$ after the $s$
photon has traversed the $\lambda/4$ plates and two slits [see their Figure 1]
does not retrocause the $s$ photon to already have "polarization $y$ before
hitting the $\lambda/4$ plates." When photon $p$ is measured with polarization
$x$, then the two particle system is in the superposition state:

\begin{center}
$i\left\vert R\right\rangle _{s1}\otimes\left\vert x\right\rangle
_{p}-i\left\vert L\right\rangle _{s2}\otimes\left\vert x\right\rangle
_{p}=\left[  i\left\vert R\right\rangle _{s1}-i\left\vert L\right\rangle
_{s2}\right]  \otimes\left\vert x\right\rangle _{p}$
\end{center}

\noindent which means that the $s$ photon is \textit{still} in the
slit-superposition state: $i\left\vert R\right\rangle _{s1}-i\left\vert
L\right\rangle _{s2}$. Then only with the measurement of the circular
polarization states $L$ or $R$ at $D_{s}$ do we have the collapse to (the
evolved version of) one of the slit eigenstates $s1$ or $s2$ (in their
notation). It is an instance of the separation fallacy to infer "the passage
of $s$ through slit 1" or "slit 2", i.e., $s1$ or $s2$, instead of the photon
$s$ being in the tagged superposition state $\left\vert \Psi\right\rangle $
after traversing the slits.

Let us take a new polarization space basis of $\left\vert +\right\rangle
=+45^{\circ}$ to the $x$-axis and $\left\vert -\right\rangle =-45^{\circ}$ to
the $x$-axis. Then the overall state can be rewritten in terms of this basis
as (see original paper for the details):

\begin{center}
$\left\vert \Psi\right\rangle =\frac{1}{2}\left[  \left(  \left\vert
+\right\rangle _{s1}-i\left\vert +\right\rangle _{s2}\right)  \otimes
\left\vert +\right\rangle _{p}+i\left(  \left\vert -\right\rangle
_{s1}+i\left\vert -\right\rangle _{s2}\right)  \otimes\left\vert
-\right\rangle _{p}\right]  $.%

\begin{center}
\includegraphics[
natheight=2.787500in,
natwidth=2.666800in,
height=2.8857in,
width=2.7607in
]%
{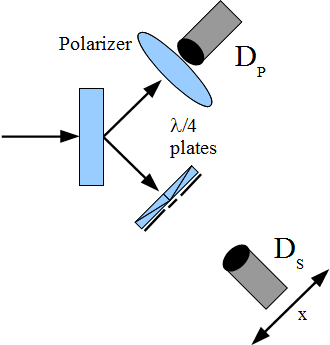}%
\\
Figure 14: Setup with $D_{p}$ polarizer
\end{center}

\end{center}

Then a $\left\vert +\right\rangle $ polarizer or a $\left\vert -\right\rangle
$ polarizer is inserted in front of $D_{p}$ to select $\left\vert
+\right\rangle _{p}$ or $\left\vert -\right\rangle _{p}$ respectively. In the
first case, this reduces the overall state $\left\vert \Psi\right\rangle $ to
$\left\vert +\right\rangle _{s1}-i\left\vert +\right\rangle _{s2}$ which
exhibits an interference pattern, and similarly for the $\left\vert
-\right\rangle _{p}$ selection. This is misleadingly said to "erase" the
so-called "which-slot information" so that the interference pattern is restored.

The first thing to notice is that two complementary interferences patterns,
called "fringes" and "antifringes," are being selected. Their sum is the
no-interference pattern obtained before inserting the polarizer. The polarizer
simply selects one of the interference patterns out of the mush of their
merged non-interference pattern. Thus instead of "erasing which-slit
information," it selects one of two interference patterns out of the
both-patterns mush.

Even though the polarizer may be inserted after the $s$-photon has traversed
the two slits, there is no retrocausation of the photon going though both
slits or only one slit as previously explained.

One might also notice that the entangled $p$-photon plays little real role in
this setup (as opposed to the "delayed erasure" setup considered next).
Instead of inserting the $\left\vert +\right\rangle $ or $\left\vert
-\right\rangle $ polarizer in front of $D_{p}$, insert it in front of $D_{s}$
and it would have the same effect of selecting $\left\vert +\right\rangle
_{s1}-i\left\vert +\right\rangle _{s2}$ or $\left\vert -\right\rangle
_{s1}+i\left\vert -\right\rangle _{s2}$ each of which exhibits interference.
Then it is very close to the one-photon eraser experiment of the last section.

\section{Delayed quantum eraser}

If the upper arm is extended so the $D_{p}$ detector is triggered last
("delayed erasure"), the same results are obtained. The entangled state is
then collapsed at $D_{s}$. A coincidence counter (not pictured) is used to
correlate the hits at $D_{s}$ with the hits at $D_{p}$ for each fixed
polarizer setting, and the same interference pattern is obtained.%

\begin{center}
\includegraphics[
natheight=2.489000in,
natwidth=2.520800in,
height=2.5779in,
width=2.6123in
]%
{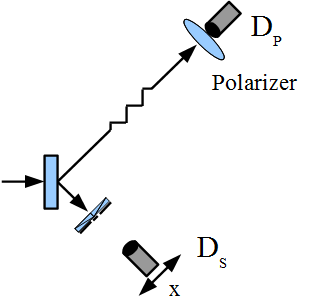}%
\\
Figure 15: Setup for "delayed erasure"
\end{center}

The interesting point is that the $D_{p}$ detections could be years after the
$D_{s}$ hits in this delayed erasure setup. If the $D_{p}$ polarizer is set at
$\left\vert +\right\rangle _{p}$, then out of the mush of hits at $D_{s}$
obtained years before, the coincidence counter will pick out the ones from
$\left\vert +\right\rangle _{s1}-i\left\vert +\right\rangle _{s2}$ which will
show interference.

Again, the years-later $D_{p}$ detections do not retrocause anything at
$D_{s}$, e.g., do not "erase which-way information" years after the $D_{s}$
hits are recorded (in spite of the "delayed erasure" talk). They only pick
(via the coincidence counter) one or the other interference pattern out of the
years-earlier mush of hits at $D_{s}$.

\begin{quotation}
"We must conclude, therefore, that the loss of distinguishability is but a
side effect, and that the essential feature of quantum erasure is the
post-selection of subensembles with maximal fringe visibility." \cite[p.
79]{Kwiat:erase}
\end{quotation}

The same sort of analysis could be made of the delayed choice quantum eraser
experiment described in the paper by Kim et al. \& Scully \cite{Kim:eraser}.
Brian Greene \cite[pp. 194-199]{Greene:fabric} gives a good informal analysis
of the Kim et al. \& Scully experiment which avoids the separation fallacy and
thus avoids any implication of retrocausality.

\end{document}